# Nanoscale correlated disorder in out-of-equilibrium myelin ultrastructure

Gaetano Campi[1], Michael Di Gioacchino[1,2,3], Nicola Poccia[4], Alessandro Ricci[2], Manfred Burghammer[5], Gabriele Ciasca[6], Antonio Bianconi[1,2,7]

[1] Institute of Crystallography, CNR, via Salaria, Km 29.300, 00015 Monterotondo Roma, Italy
[2] Rome International Center for Materials Science Superstripes (RICMASS), Via dei Sabelli 119A, 00185 Roma, Italy
[3] Department of Science, Nanoscience section, Roma Tre University, Via della Vasca Navale 84, 00146 Roma, Italy
[4] Department of Physics, Harvard University, Cambridge, Massachusetts 02138, USA∗
[5] European Synchrotron Radiation Facility, 6 Rue Jules Horowitz, BP220, 38043 Grenoble Cedex, France
[6] Physics Institute, Catholic University of Sacred Heart, Largo F. Vito 1,00168 Rome, Italy
[7] National Research Nuclear University MEPhI (Moscow Engineering Physics Institute) 115409 Moscow Kashirskoe shosse 31 Russia

**Abstract**

Ultrastructural fluctuations at nanoscale are fundamental to assess properties and functionalities of advanced out-of-equilibrium materials. We have taken myelin as a model of supramolecular assembly in out-of-equilibrium living matter. Myelin sheath is a simple stable multi-lamellar structure of high relevance and impact in biomedicine. Although it is known that myelin has a quasi-crystalline ultrastructure there is no information on its fluctuations at nanoscale in different states due to limitations of the available standard techniques. To overcome these limitations, we have used Scanning micro X-ray Diffraction, which is a non-invasive probe of both reciprocal and real space to visualize statistical fluctuations of myelin order of the sciatic nerve of Xenopus Laevis. The results show that the ultrastructure period of the myelin is stabilized by *large anti-correlated* fluctuations at nanoscale, between hydrophobic and hydrophilic layers. The ratio between the total thickness of hydrophilic and hydrophobic layers defines the conformational parameter, which describes the different states of myelin. Our key result is that myelin in its *out-of-equilibrium* functional state fluctuates point-to-point between different conformations showing a correlated disorder described by a Levy distribution. As the system approaches the thermodynamic equilibrium in an aged state the disorder loses its correlation degree and the structural fluctuation distribution changes to Gaussian. In a denatured state at low pH, it changes to a completely disordered stage. Our results clarify also the degradation mechanism in biological systems by associating these states with variation of the ultrastructural dynamic fluctuations at nanoscale.

**Introduction**

A hot topic for material scientists today is the design of chemical systems in quasi stationary non-equilibrium states.[1] In fact it is known that chemical systems can acquire novel functionalities such as for example chiral symmetry breaking, bistability and ultimately life,[2] when kept away from equilibrium. In this context, supramolecular self-assembly is of high interest. It involves multiple weak intermolecular interactions between structural units leading to multiple conformational states with similar energy separated by small potential energy barriers. The interchange among the different structural conformations moves the systems along the complex potential energy landscape, which determines their functionality. Thus, the fluctuations between the nearly isoenergetic conformations constitute a key point to be investigated for a deeper understanding of out-of-equilibrium matter, that occurs in biomaterials,[3] plants,[4] nanotechnology[5] and proteins.[6] In particular, the spatial correlation degree of these fluctuations defines and characterizes the system thermodynamic state related to its function.

It has been proposed in recent theories that the out-of-equilibrium state of living biological matter is close to a critical point.[7-10] The most significant indicator for proximity to a critical point is critical opalescence where the size of the regions of different phases begin to fluctuate, in a correlated way, over increasingly large length scales.[11] The spatial correlations are described by universal power laws and scaling functions in soft, biological as well as in hard matter.[12-15] Recently Kaufman group has proposed quantum criticality at origin of life[10] predicting biological matter fluctuations as in a system in proximity to a quantum critical point for metal-insulator transition (MIT). Power law fluctuations in MIT quantum criticality have been well characterized in high temperature superconductors.[16-19]

Myelin[20,21] can be considered the simplest example of a biological ultrastructure since it shows a periodic multilayer structure rich in phospholipid membranes with few active proteins[22-28] where the competition between fundamental biological interactions[29] determines its quasi steady state out-of-equilibrium. The compact myelin sheath is an elaborated multi-layered membrane wrapping selected axons. Its main role is the control of the propagation speed of action potentials in saltatory conduction, facilitating nerve signal transmission.[20,21] The myelin ultrastructure is usually described as a highly-ordered liquid crystal[22] made of a structural unit constituted by the stacking of the four layers: *i*) lipidic membrane (lipid polar group, *lpg*) *ii*) cytoplasmatic apposition, *cyt*, *iii*) a second lipidic membrane, *lpg*, iv) extracellular apposition, *ext*.[27,30] The structural unit appears to be stable, as probed by using a wide variety of standard experimental approaches[25,30-40] giving information only on the average structure, and thus inadequate for the visualization of out-of-equilibrium structural fluctuations, which require highly spatially resolved probes. Besides, on account of the advanced features of the latest generation synchrotron sources and fast acquisition methods, nowadays powerful non-invasive high spatial resolution techniques are available for this purpose.[41,42] In our experimental approach, we have used Scanning micro X-ray Diffraction (SµXRD) to probe the k-space (reciprocal space) order with high resolution in the real space. In this way we have mapped spatial distribution of myelin ultrastructural fluctuations *e.g.* the fluctuations of myelin sublayers thickness. In a second step, we have applied statistical physics tools to the collected diffraction data to unveil the "statistical distributions of the fluctuating structural order". This approach has recently been used to carry out heterogeneity investigation of complex matter in several systems of interest in different fields from biomedicine to material science.[26,27,41-43]

We have applied this approach to investigate the myelin in the sciatic nerves of Xenopus Leavis frogs, which is a Peripheral Nervous System (PNS) representative. We have measured the spatial fluctuations of the four nanoscale layers thickness at thousands of discrete locations in functional, out-of-equilibrium freshly extracted nerves associated with the *in vivo* state,[33] named *fresh* samples. Afterwards, in order to investigate the system degradation towards the thermodynamic equilibrium we quantified the structural statistical fluctuations of myelin in aged nerves after extraction, named *unfresh* and denatured nerves in acidic pH solution, named *denatured*. We have used the ratio between the thickness of the hydrophilic and hydrophobic layers called hydrophilic-hydrophobic conformational parameter (HhCF) to characterize the myelin local state. We discovered that the quasi-crystalline periodicity of the myelin lattice in the functional state is maintained thanks to large anti-correlated intrinsic fluctuations between the hydrophobic layers and hydrophilic layers. In *fresh* out-of-equilibrium samples, the probability distribution of HhCF shows a fat tail, here described by a Levy distribution,[44-49] as expected for biological matter near a quantum critical point.[10-19,50,51] Leaving the functional out-of-equilibrium state, towards the thermodynamic equilibrium the myelin degrades. In the aged state, we have observed the freezing ultrastructure fluctuations leading to a narrow Gaussian distribution of HhCF, while in denatured state,[35] at low pH, we observed a disordered state indicated by the larger uncorrelated fluctuations. These results support the hypothesis that the functional *fresh* nerve is a system in a non-equilibrium steady state tuned close to a critical point.[7-10]

# Results and discussion

The ultrastructure of myelin is shown schematically in **Figure 1a**. The repetition of the structural unit, made by the stacking of *i)* cytoplasmatic (*cyt*), *ii)* lipidic (*lpg*), *iii)* extracellular (*ext*) and *iv)* another lipidic (*lpg*) [27,32] layers, gives rise to the Bragg peaks well known in literature.[20,26,30,32] The XRD profiles with the indicated Bragg peaks of order 2, 3, 4 and 5, measured in 1 μm² area, of *fresh*, *unfresh* and *denatured* samples are shown in **Figure 1b**. The individual thickness of each layer, named $d_\lambda$, $d_{cyt}$, $d_{lpg}$, $d_{ext}$, have been extracted from electron density (ED) profiles, for the three samples, *fresh*, *unfresh* and *denatured*, shown in **Figure 1c**. These ED profiles have been computed by Fourier analysis of the XRD profiles, as described in detail in Materials and Methods. Different features in XRD and ED peaks shapes are quite evident; this necessitates a spatially resolved study jointly to the application of spatial statistical tools to describe the ultrastructural fluctuations in this system.

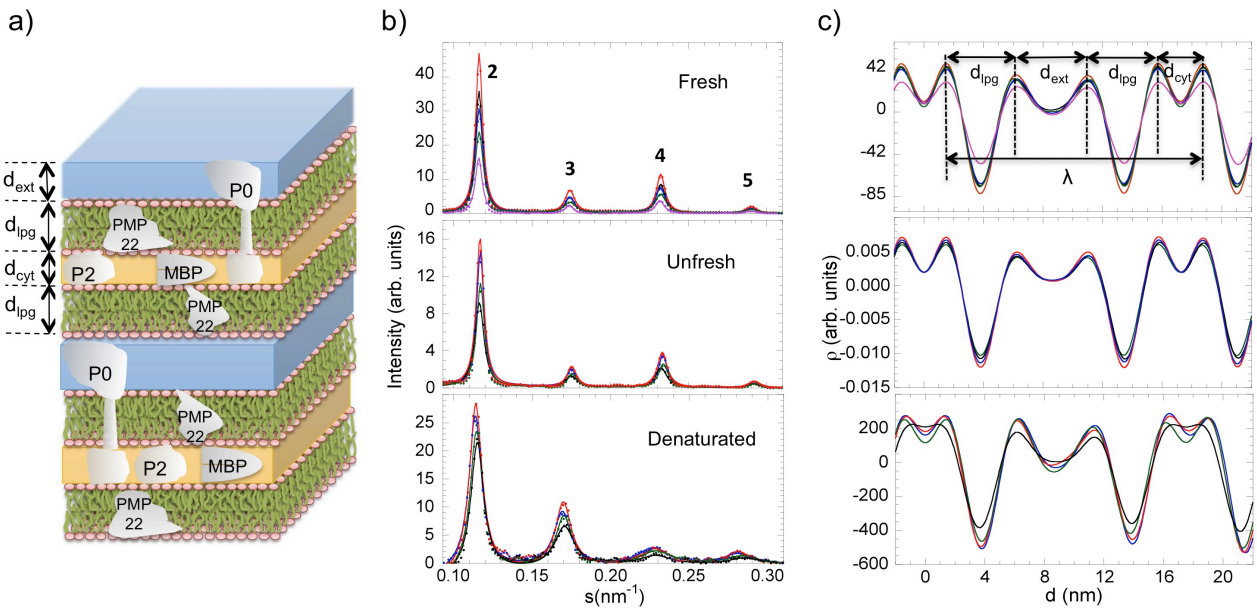

**Figure 1: a)** Pictorial view of the protein depleted membrane layers made of polar lipid groups, *lpg*, with thickness $d_{lpg}$, intercalated by two hydrophilic layers: the cytoplasmic apposition, *cyt*, and the extracellular apposition, *ext*, with thickness $d_{cyt}$ and $d_{ext}$. respectively. The location of myelin proteins PMP22, P0, P2 and MBP is indicated. **b)** The diffracted intensity is plotted against the reciprocal distance, s, and the Bragg reflections of order 2, 3, 4, 5 are indicated for each sample. **c)** The Electron density distribution has been computed from the diffraction patterns, shown in Figure 1b, using Fourier analysis. The curve presents minima and maxima, with a period $d_\lambda = 2d_{lpg} + d_{ext} + d_{cyt}$, where $d_{lpg}$, $d_{ext}$ e $d_{cyt}$ refer to the thickness of lipidic (*lpg*), extracellular (*ext*), cytoplasmatic (*cyt*) zones respectively. The Electron density distribution for unfresh and denatured sample have been obtained in the same way of the other fresh one, shown in **b)**.

To this purpose, we have scanned the different nerves in 300x125 $\mu m^2$ central Regions of Interest (ROI) selected along the vertical axis of the nerves. First, we have built the maps of the thickness of structural unit period, $d_\lambda$, and of each sub-layer $d_{cyt}$, $d_{lpg}$, $d_{ext}$, in the *fresh* samples, as shown in **Figure 2a, 2c, 2e** and **2g**, respectively.

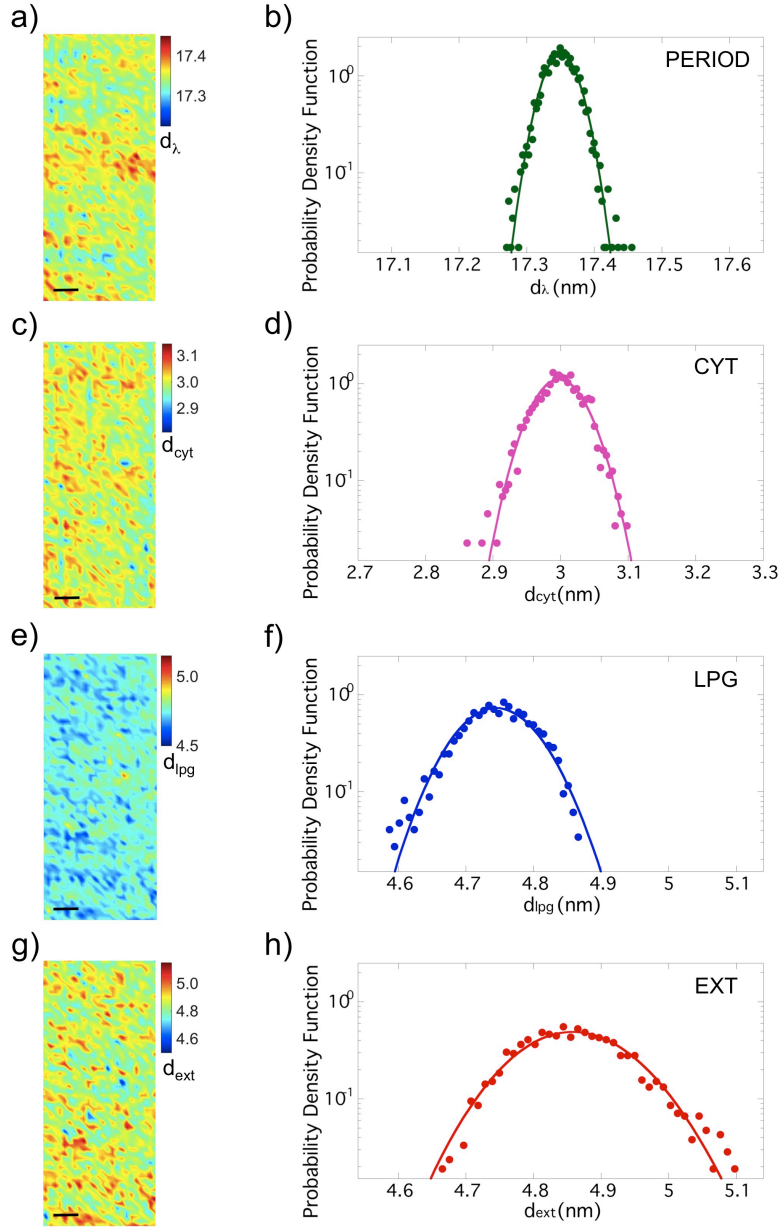

**Figure 2:** Maps (300x125 $\mu m^2$) of $d_\lambda$, $d_{cyt}$ $d_{lpg}$ and $d_{ext}$ in **a)**, **c)**, **e)**, and **g)** respectively. The probability density functions (PDFs) of these thickness maps are shown in **b)**, **d)**, **f)**, **h)**. We notice the larger fluctuations of $d_{cyt}$, $d_{ext}$ and $d_{lpg}$ given by $\sigma(d_{cyt})$, $\sigma(d_{ext})$ and $\sigma(d_{lpg})$ respectively, with respect to $\sigma(d_\lambda)$ fluctuations of the period $d_\lambda$ (see Table 1).

Then, we calculated the Probability Density Function (PDF) of the same quantities (see **Figure 2b, 2d, 2f** and **2h**). The myelin *ultrastructural spatial fluctuations* have been defined by the relative standard deviation, $RSD_i=\sigma(d_i)/\mu(d_i)$ on the whole ROI, where $\mu(d_i)$ and $\sigma(d_i)$ are the mean value and the standard deviation of the probability distribution for each thickness, $d_i$, with $i = \lambda, cyt, ext$ and $lpg$.

Let us consider the thickness of the periodic structural unit, $d_\lambda$, (Fig. 2a, 2b). Its average value is found to be $\mu(d_\lambda) = 17.350$ nm, in agreement with previous works.[26,30] Its spatial fluctuation, $RSD_\lambda=\sigma(d_\lambda)/\mu(d_\lambda)$, results to be about 0.14%, quite smaller than the fluctuations for the other individual sublayers, having RSD between 1.1-1.7% as reported in **Table 1**. Here we meet the "paradox" of myelin *quasi-crystallinity* due to the smaller spatial fluctuation of period in comparison with the larger sublayers' fluctuations.

|  | μ(nm) | | | σ(nm) | | | RSD (%) | | |
|---|---|---|---|---|---|---|---|---|---|
| samples | *fresh* | *unfresh* | *pH=5* | *fresh* | *unfresh* | *pH=5* | *fresh* | *unfresh* | *pH=5* |
| $d_{cyt}$ | 2.999 | 2.957 | 2.832 | 0.036 | 0.028 | 0.237 | 1.20 | 0.95 | 8.37 |
| $d_{ext}$ | 4.860 | 4.797 | 5.597 | 0.082 | 0.049 | 0.292 | 1.68 | 1.03 | 5.22 |
| $d_{lpg}$ | 4.746 | 4.798 | 4.669 | 0.055 | 0.034 | 0.256 | 1.15 | 0.71 | 5.49 |
| $d_\lambda$ | 17.350 | 17.350 | 17.960 | 0.024 | 0.021 | 0.121 | 0.14 | 0.12 | 0.68 |

**Table 1**: Statistical analysis of the of the thickness $d_{cyt}$, $d_{ext}$, $d_{lpg}$ and $d_\lambda$, in the fresh, unfresh and denatured with pH=5 samples, from the electron density profiles for Xenopus Laevis sciatic nerves. Here we report mean values, μ, standard deviations, σ, and Relative Standard Deviation RSD = σ/μ.

| $c_{i-j}$ | $d_{cyt}$ | | | $d_{ext}$ | | | $d_{lpg}$ | | |
|---|---|---|---|---|---|---|---|---|---|
| samples | *fresh* | *unfresh* | *pH=5* | *fresh* | *unfresh* | *pH=5* | *fresh* | *unfresh* | *pH=5* |
| $d_{cyt}$ | 1 | 1 | 1 | 0.4452 | 0.3426 | 0.7283 | -0.6406 | -0.5293 | -0.895 |
| $d_{ext}$ | | | | 1 | 1 | 1 | -0.9474 | -0.8290 | -0.925 |
| $d_{lpg}$ | | | | | | | 1 | 1 | 1 |

**Table 2**: Correlation coefficients $c_{i-j}$ between maps of the spatial fluctuations between the different layers i, j = *cyt*, *ext* and *lpg* in the fresh, unfresh and pH=5 samples. Negatively correlated coefficients are indicated by grey cells. These correlations are well explained by Figure 6 for each sample.

To shed light on this paradox, we have studied the correlations between the spatial fluctuations $RSD_{lpg}$, $RSD_{cyt}$ and $RSD_{ext}$. In order to do this, we first calculated the Pearson's correlation coefficients, $c_{i-j}$, between maps of cytosolic apposition (Fig. 2c), extracellular apposition (Fig. 2g), lipidic membrane (Fig. 2e) (see **Table 2**). We observed that the spatial fluctuations $RSD_{cyt}$ are smaller than $RSD_{ext}$ but $d_{cyt}$ and $d_{ext}$ wave together in the same direction, being $c_{cyt-ext}$ positive. On the other hand, fluctuations $RSD_{lpg}$ go in the opposite direction since $c_{lpg-ext}$ and $c_{cyt-lpg}$ are negative. This means that the lipid membrane fluctuations are strongly anti-correlated with both cytosolic and extracellular layer fluctuations. It is natural to associate this anti-correlation with the anti-correlated dynamics of the myelin proteins.[25,52,53,54,55] In order to describe the point-to-point spatial structural fluctuations in myelin we have introduced the hydrophilic-hydrophobic conformational parameter (HhCF) described by the ratio $\xi$ between hydrophilic and hydrophobic layers:

$$\xi=(d_{ext}+d_{cyt})/2d_{lpg} \qquad (1)$$

A typical map of $\xi$ measured on the *fresh* samples is shown in **Figure 3a**. The red spots represent areas where hydrophilic layers are larger, while in the blue spots they become smaller and the thickness of hydrophobic layers increase. The fact that the $\xi$ is always less than 0.9 demonstrates that myelin has a high lipid content.[20]

The PDF of $\xi$ shows a skewed line shape modelled by using Levy distributions[44] (see **Figure 3b**). The Levy distributions provide a statistical description of complex signals deviating from normal behaviour and in recent years have found increasing interest in several applications in diverse fields.[45-49] The characteristic function of Levy probability distribution is defined by four parameters: stability index α, skewness parameter β, scale parameter γ taking into account the width of the distribution, and location parameter δ with varying ranges of 0<α≤2, -1≤β≤1, γ>0 and δ real. We stress the fact that the closed form expressions of density and distribution functions of Levy distributions are not available except for few particular cases such as the well-known Gaussian distribution where the stability parameter α is 2. Here we have used basic functions in the numerical evaluation of these parameters and goodness of data fitting as described by Liang and Chen.[56] The Levy fitting curve, indicated by the continuous line in Figure 3b, gives a stability index of 1.78±0.02 (< 2), a location of 0.82±0.01, a skewness of 1.00±0.05 and a scale parameter γ of 0.014±0.002.

The ultrastructural dynamic fluctuations and the myelin quasi-crystallinity paradox can be seen in a compact and clear way in **Figure 3c.** Here we show the scatter plot of the relative variations of the layers, $\Delta d_{ext}=d_{ext}-d^0_{ext}$, $\Delta d_{lpg}= d_{lpg}-d^0_{lpg}$, $\Delta d_{cyt}= d_{cyt}-d^0_{cyt}$ and the period, $d_\lambda$, as a function of $\xi$, where $d^0_{ext}$, $d^0_{lpg}$, and $d^0_{cyt}$ are constant values given by 2.2, 3.7 and 2.2 nm, respectively, which are the thicknesses of the membrane units in the protein-free myelin sheath.[25] We observed that the almost total independence of the period from $\xi$ is due to the positive correlations between the fluctuations of hydrophilic layers ($\Delta d_{ext}$ and $\Delta d_{cyt}$) which increase with increasing $\xi$, and their anti-correlation with hydrophobic fluctuations, $\Delta d_{lpg}$, which decrease with the increase of $\xi$. This clearly shows that hydrophilic fluctuations compensate the hydrophobic fluctuations to keep the period almost constant giving the apparent static crystalline-like character of the myelin ultrastructure.

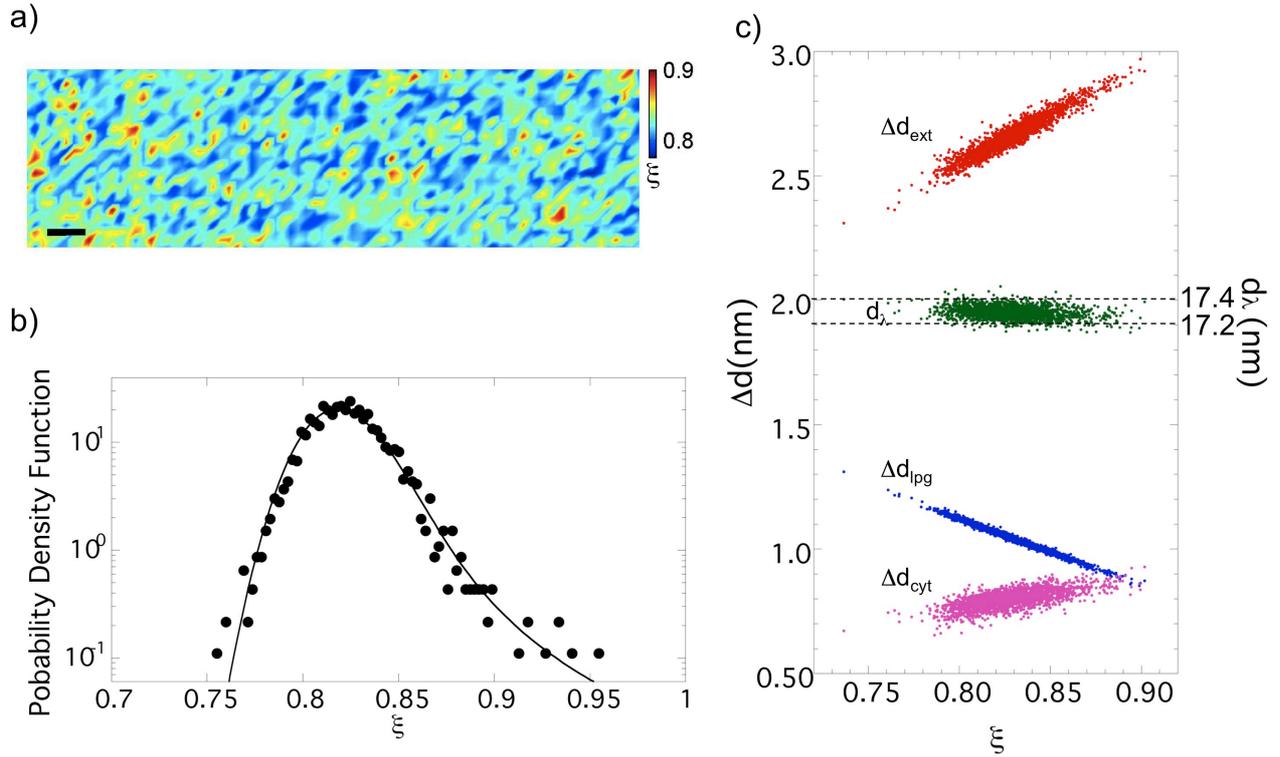

**Figure 3: a)** Map of the HhCF conformational parameter, $\xi$, in a selected central zone of the nerve (300x125 $\mu m^2$) and probability density function of $\xi$ in semi-log plot. Here we notice the fat tail of that PDF fitted with Levy distribution. **b)** Scatter plot of absolute dispersion $d_\lambda$, $\Delta d_{ext}=d_{ext}-d^0_{ext}$, $\Delta d_{lpg}=d_{lpg}-d^0_{lpg}$, $\Delta d_{cyt}=d_{cyt}-d^0_{cyt}$ as a function of $\xi$, where $d^0_{ext}=d^0_{cyt}=2.2$ nm and $d^0_{lpg}=3.7$ nm. The period, $d_\lambda$, is reported enclosed between two dotted lines at $d_\lambda=17.4$ *nm* and $d_\lambda=17.2$ nm. This plot solves the paradox of the period stability. We note the almost total independence of the period from $\xi$ due to the anti-correlated fluctuations between the hydrophilic layers (*cyt* and *ext*) and the hydrophobic one (*lpg*).

Subsequently, we proceeded to check if the described fluctuations are an intrinsic feature of the functional out-of-equilibrium state of myelin, studying the ultrastructural fluctuations in the two different following cases:

1) an aged state, named *unfresh*, where the sciatic nerve has been left for 18 hours in a Ringer solution after the dissection, without insertion of Adenosyne-Tri-Phosphate (ATP) and oxygen where the system is expected to go toward *an equilibrium state* from its functional *stationary state out-of-equilibrium*;

2) a state with acid pH (pH=5), named *denatured*, where the sciatic nerve has been left in an acid buffer solution where the system is expected to go toward *degeneration*.[35]

The ultrastructural fluctuations of the *unfresh* and *denatured* samples are shown in **Figure 4**. Typical maps and the PDF of the $d_\lambda$, $d_{cyt}$, $d_{lpg}$, $d_{ext}$ thicknesses of the *unfresh* nerve are shown respectively in **Figure 4a**, **4b**, **4c** and **4d**.

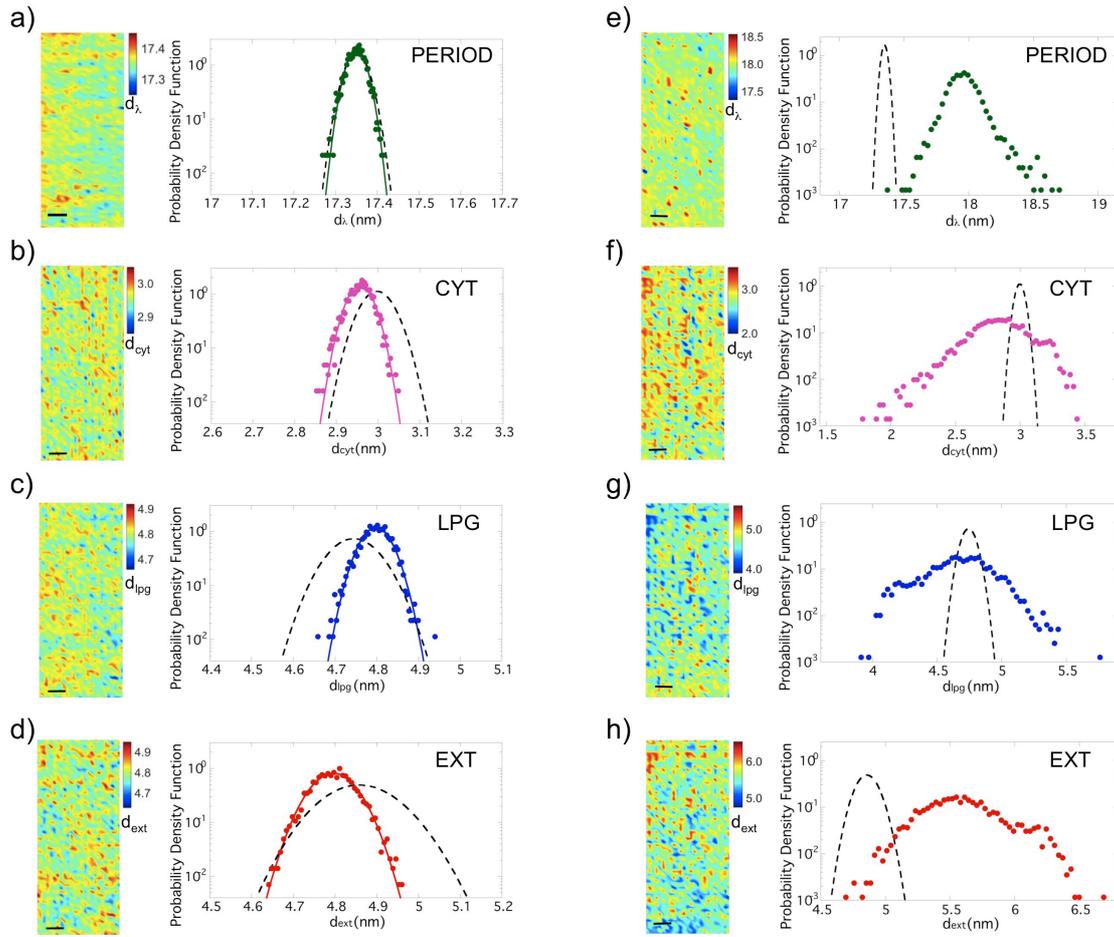

**Figure 4**: Maps (300x125 $\mu m^2$) of $d_\lambda$, $d_{cyt}$ $d_{lpg}$ and $d_{ext}$ with measured PDF in the unfresh sample, depicted in **a)**, **b)**, **c)**, **d)**, while in **e)**, **f)**, **g)**, and **h)** is represented the maps and relative measured PDFs for the denatured sample. The dashed lines represent the fresh sample to comparison with the other sample. In the unfresh sample, despite the unchanged period, in comparison with the fresh nerve, we observed reduced fluctuations of $d_{cyt}$, $d_{ext}$ and $d_{lpg}$ (see RSD values in Table 1). These narrower fluctuations are associated with the loss of correlated disorder, as can be seen by the decreasing of the Pearson's correlation coefficient for the unfresh nerve in Table 2. In the case of denaturation with pH shows bigger increasing of fluctuations (see RSD values in Table 1) and the distribution cannot be mathematically modelled, showing in the sign of degeneration stage.

The dashed lines represent the PDF of the thicknesses in the functional *fresh* nerve, for comparison. The average *unfresh* myelin period, $d_\lambda$ = 17.350 ± 0.021 nm, results to be the same as in the *fresh* state. This indicates that the myelin is not yet in any morphologically degenerate state where the period is expected to change significantly.[26,30] On the other hand, we observe significant change in the dynamics of the system. Indeed, both the standard deviations and the mean values of each individual layer thickness change in comparison with the *fresh* functional state. In particular, the spatial fluctuations decrease significantly, as we can see by comparing the RSD values in **Table 1**. We observe that the lipidic hydrophobic thickness, $d_{lpg}$, reaches smaller values down to about 4.6 nm in the *fresh* functional state in comparison with the *unfresh* state. At the same time the hydrophilic extracellular layer reaches larger values up to 5.1 nm in the *fresh* nerve while it does not exceed 5 nm in the *unfresh* state. The decreasing fluctuations in the *unfresh* nerve are accompanied by the reduction of the spatial correlations, as indicated by comparing the correlation coefficients between cytosolic, extracellular and lipidic membrane apposition in **Table 2**. Thus, although the average period in the aged *unfresh* nerve is unchanged we find *i)* decreasing amplitude of fluctuations and *ii)* decreasing spatial correlations between fluctuations. The myelin *frozen fluctuations* support the hypothesis that the aged system has reached a more static state, approaching the thermodynamic equilibrium.

Let us now consider the *denatured* sample at pH=5. The maps and the relative PDFs of the $d_\lambda$, $d_{cyt}$, $d_{lpg}$, $d_{ext}$ thicknesses are shown in **Figure 4e**, **4f**, **4g** and **4h**, respectively. The period PDF (Fig. 4e) shows a clear increase in both its mean value and in its standard deviation compared to the fresh sample. In this case $\mu(d_\lambda)$ = 17.960 and $\sigma(d_\lambda)$ = 0.121 nm give spatial fluctuations RSD($d_\lambda$) = 0.68%, much larger than fluctuations in both *fresh* and *unfresh* states. This is an expected degradation sign of the morphology, indeed, as is commonly known, the degeneration leads to an increase in the period and later breaking of the biological membrane.[30,40]

Alongside the morphologically changes, we observed also dramatic changes in the dynamics of the system. Upon inspection of Figure 4f, 4g and 4h we notice a huge increase of the range of spatial fluctuations for all sublayers, compared to the fresh sample. Indeed, now the RSD assume values including between 5.2-8.4%, (see Table 1). The PDFs assume now a non-analytic shape, sign of sample denaturation process. In particular, focusing our attention on cytosolic apposition PDF (Fig. 4f) we can see that there is a long PDF tail at low values of $d_{cyt}$, which reaches the range of the minimum possible *cyt* thickness;[25] on the contrary a sharp drop of PDF is visible at high values of $d_{cyt}$. In order to check the correlation between each sub-layer in the denatured sample, we direct our attention to Pearson's correlation coefficients, $c_{i-j}$, between maps of cytosolic apposition (Fig. 4f), extracellular apposition (Fig. 4h), lipidic membrane (Fig. 4g). We notice a great enhancement of the coefficient $c_{cyt-ext}$ and a great decreasing of the coefficient $c_{cyt-lpg}$, that shows the correlation and the anti-correlation between the cytosolic apposition and the other two sub-layers, in comparison with *fresh* state (see **Table 2**). At any rate, the bigger coefficient is the anti-correlation $c_{ext-lpg}$, which is however just a little lower than the *fresh* state. Therefore, in general, the sign of coefficient $c_{i,j}$ between sublayers in each sample are maintained, suggesting that these correlations are a structural property more than functional. The structural anti-correlation occurs between $d_{lpg}$ and $d_{ext}$. This is important because it means that the degradation leads to loss of fluctuations compensation between

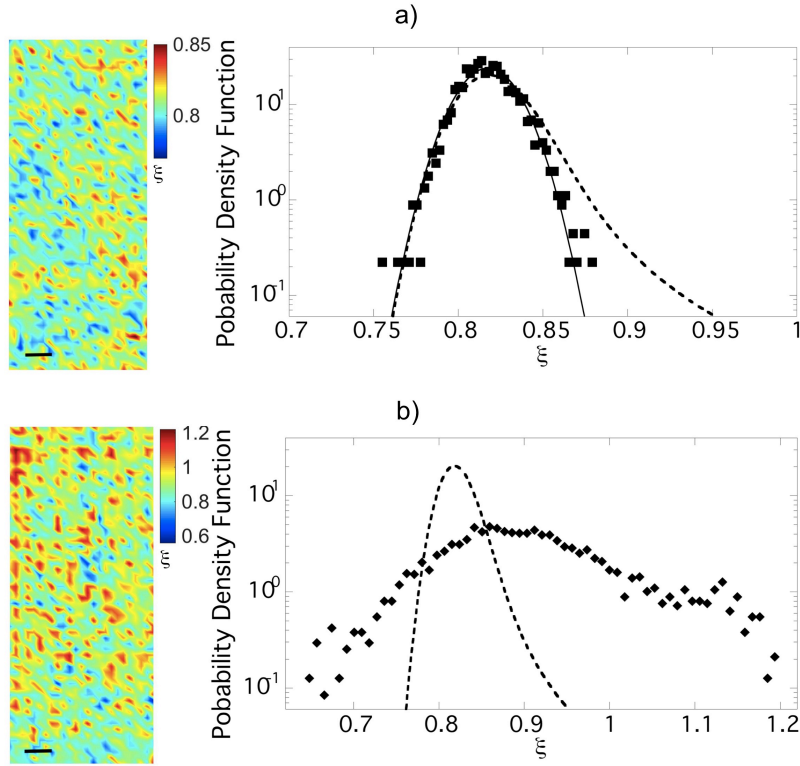

**Figure 5: a)** Map of the conformational parameter, $\xi$, in a selected central zone of the unfresh nerve (300x125 $\mu m^2$) and probability density function of $\xi$ in the unfresh sample (squared) in semi-log plot. The Levy PDF curve found in the fresh sample (dashed line) is also reported, for comparison. We notice the loosing of the fat tail in the distribution assuming a Gaussian profile in the unfresh sample. **b)** Maps (300x125 $\mu m^2$) of the conformational parameter $\xi$ at the pH=5 sample and his probability density function in semi-log plot. The dashed line indicated the fit of the fresh sample with Levy distribution fat tail. The denatured PDFs isn't fitted because not easily mathematically descript and change shape. The parameter $\xi$ is useful to characterize the overall myelin state uniquely. The map scale bar corresponds to 25 $\mu m$.

Our main result can be illustrated by comparing the PDF of the conformational parameter, *ξ* maps, in the *fresh*, *unfresh* and *denatured* samples, shown in **Figure 5a** and **Figure 5b** respectively. The dashed lines represent the PDF of *ξ* in the functional *fresh* nerve. Focusing our attention on Fig. 5a, we note that the PDF becomes Gaussian in the *unfresh* nerve, losing the fat tail and the correlated disorder at the nanoscale, modelled by Levy flight in the *fresh* sample. Indeed, the stability parameter here becomes equal to 2.00±0.02. Furthermore, there is a decrease of both the location parameter δ = 0.81±0.01 and of the scale parameter γ = 0.013±0.002 compared to the fresh sample, which identifies the narrowing of hydrophilic and the increasing of hydrophobic layers.

Regarding the pH=5 sample, as for the individual sub-layers, the PDF of *ξ* does not follow any clear specific trends, showing the sign of disordering in the denaturation process. There is a substantial expansion of the distribution, associated with an increase in mean value and fluctuation around it. It results to be $\xi = 0.9 \pm 0.3$, that shows a great denaturation because it allows values of $\xi$ larger than 1, changing the lipid rich myelin structure in standard physiological form to the lipid poor myelin with a greater amount of hydrophobic part in degenerated state.

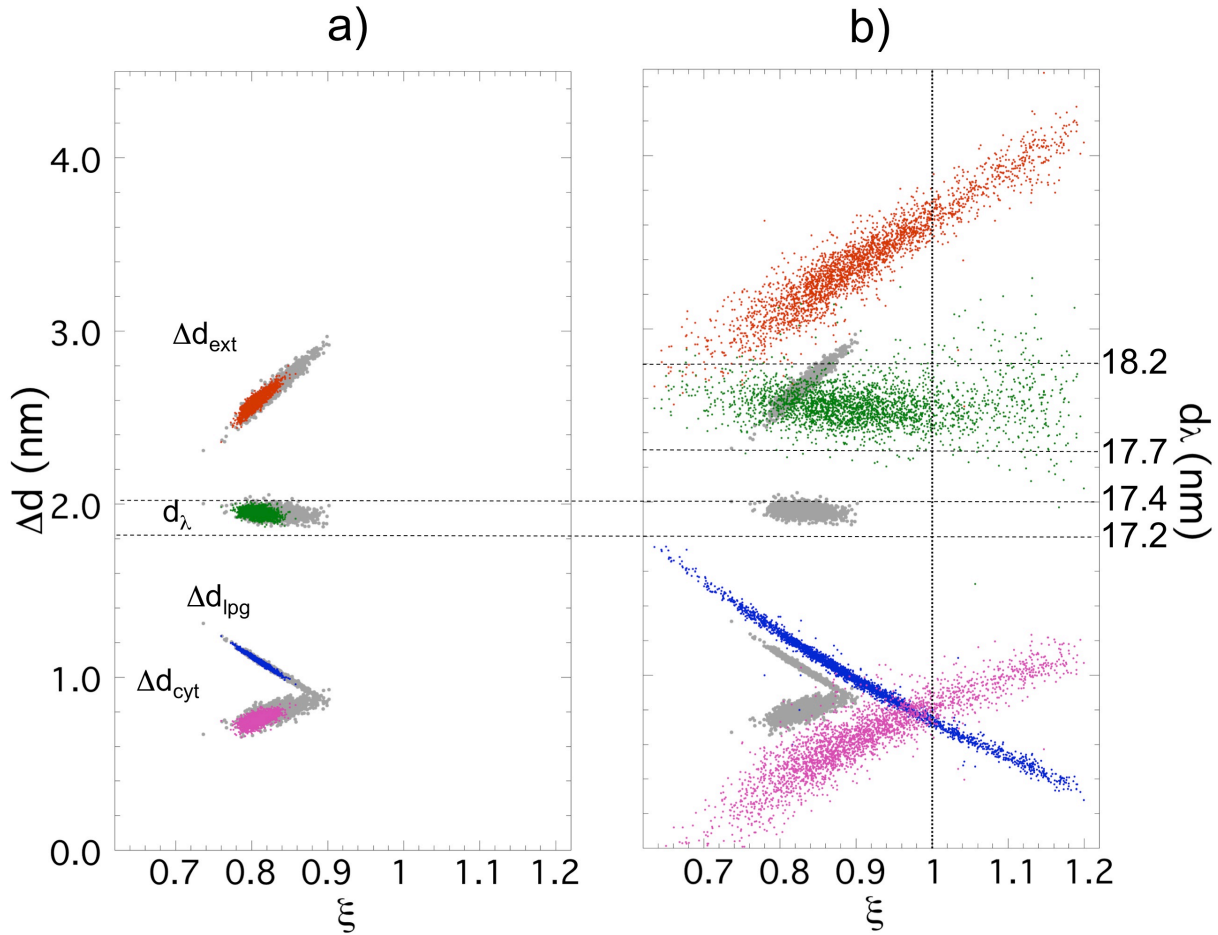

**Figure 6: a)** Scatter plot of the period $d_\lambda$ and the dispersions $\Delta d_{ext}$, $\Delta d_{lpg}$, $\Delta d_{cyt}$ (defined in Fig. 4) as a function of $\xi$, for unfresh sample. The period fluctuations are enclosed between two dotted lines at $d_\lambda=17.4$ *nm* and $d_\lambda=17.2$ nm. We notice the loss of fluctuations of each layer and the maintained stability of the period. The grey full circles representing the fluctuations found in the fresh nerve are reported for comparison. **b)** Scatter plot of absolute variation $\Delta d_{ext}$, $\Delta d_{lpg}$, $\Delta d_{cyt}$ and $d_\lambda$ as a function of $\xi$, for denatured sample. The period is reported on right axis between two dotted lines at $d_\lambda=18.2$ *nm* and $d_\lambda=17.7$ *nm*. The grey full circles representing the fluctuations found in the fresh nerve are reported for comparison. The period results a little bit increased, while the fluctuations have a great increasing changing the pH. The fluctuations of the hydrophilic layers (*cyt* and *ext*) and the hydrophobic one (*lpg*) are anti-correlated, *i.e.* the denatured sample maintains the anti-correlation of the functional state. But, in b), there is an across between the membrane dispersion and cytosolic apposition dispersion near $\xi=1$, that reports the passage to a degraded state, where the period shows a large disorder indicated by the larger spread of period values for $\xi>1$.

Finally, the ultrastructural fluctuations in these two non-functional states can be summarized by visualizing the relative variations of the layers, $\Delta d_{ext}$, $\Delta d_{lpg}$, $\Delta d_{cyt}$ and the period, $d_\lambda$, as a function of $\xi$. The changes in the aged sample are visualized in **Figure 6a,** where we have compared the relative variations in *unfresh* and in *fresh* sample. Despite the maintained stability of the period, the aged nerve is characterized by: *i*) decreasing fluctuations and *ii*) decreasing spatial correlations between fluctuations. This means that the aged system acquires rigidity and order, losing correlated disorder. This is quite intriguing since it tells us that in the living system the functionality is associated to the correlated disorder while the aged state shows the tendency towards a *frozen-like* state expected at equilibrium with major rigidity and order.

Upon inspection of **Figure 6b**, the *denatured* pH=5 sample shows a huge increase of fluctuations of all layers thicknesses. It is worth noting two points. First, the anti-correlative dispersions between the hydrophilic and hydrophobic layers of the structure is maintained. Second, $\xi$ reaches extreme values of 1.22 and 0.66 indicating the coexistence of very poor and very rich lipid phases, respectively. Near $\xi = 1$ we observe the intersection among $d_{lpg}$ and $d_{cyt}$ dispersions, signalled by a dashed line. This marks that the functional state $0.7 < \xi < 0.9$ coexists in the degraded states with two class of domains: one lipid-rich region that characterised by low HhCF values ($\xi < 0.7$) and another lipid-poor region characterised by high HhCF values ($\xi > 0.9$). In the lipid-poor region, the period dispersion shows a larger disorder, indicated by the larger spread of period values of about the 20%, in comparison with functional and lipid-rich domains.

As a final point, we propose a brief discussion of our achievements. First of all, it is known that, the myelin sublayers are kept together by Van der Waals (induced dipole - induced dipole) forces which ensure the integrity and stability of myelin structure.[24] In particular, the fluctuations of each sub-layer are due to its protein composition. Indeed, the active dynamics of these proteins cause the various contractions and expansions of each sub-layers.

With aging and the denaturation with pH, the myelin loses this dynamic functional ability.

The results show a loss of protein dynamic fluctuations [25,53,54,55] in the *unfresh* sample, as expected, owing to the decreasing ATP content. In fact ATP provides the energy to the myelin, controlling its functionality.[57,58]

For the *denatured* sample, the acid pH (pH=5) rebalances the Van der Waals interactions, which manifests in the variation of the sub-layers mean values ($d_{cyt}$, $d_{lpg}$, $d_{ext}$). We remark that the mean value of the hydrophobic layers remains almost constant, in comparison with the *fresh* state. On the contrary, the hydrophilic layers show large changes of their mean values: the cytosolic apposition narrows and the extracellular apposition increase, in agreement with its behaviour in the swollen nerve state,[34] as reported in Table 1.

**Conclusions**

We have obtained information on spatial nanoscale fluctuations of out-of-equilibrium PNS myelin ultrastructure of the sciatic nerve of frog Xenopus leavis using non-invasive SµXRD imaging. This technique allowed us to map the nanometric fluctuations of the myelin subcomponents: the cytoplasmatic, lipidic and extracellular layers. From statistical analysis we got the following main results: *i*) the *quasi-crystalline* stationary periodicity of the myelin lattice is due to large intrinsic anti-correlated fluctuations of hydrophobic and hydrophilic sublayers, controlled by dynamics of their proteins; *ii*) the local structure of myelin is described by the hydrophobic-hydrophilic conformational parameter HhCF given by the ratio $\xi$ between of the sum of the thickness of two hydrophilic layers (extracellular and cytoplasmatic sublayers) and the sum of the two hydrophobic lipidic membranes. We have found that the myelin structural fluctuations are characterized by regions of high hydrophobic content (small HhCF values) with rare events of more hydrophilic regions (large HhCF values) in the functional myelin, fresh state. In this sample, the hydrophobic-hydrophilic conformational parameter HhCF follows a Levy distribution with a fat-tail indicating the correlated disorder in the myelin functional biological state which we associate with the quasi stationary state in non-equilibrium tuned close to a critical point.[7-10] This is the key result of this work. This assignment is supported by the experimental data for the aged phase where the structural fluctuations are frozen, losing the correlated disorder, and the HhCF distribution changes to a narrow Gaussian centred around $\xi=0.8$ which is peculiar of a static state approaching equilibrium. Moreover, the Levy distribution disappears also in the denatured state, reached at low pH, where myelin supramolecular structure at nanoscale shows an incoherent disorder. We observe both domains with a large increase of the thickness of the extracellular hydrophilic layers with very large values $\xi >1$ at one side and on the other extreme there are domains where $\xi$ approaches the minimum possible value ~ 0.595 given by the attraction between hydrophobic layers separated by few water layers without proteins,[25] related also with denaturation of the protein structure.

The present results exploit the SµXRD technique to characterize supramolecular chemical structures in quasi stationary state out of equilibrium. Supramolecular assembly is a robust, rapid and spontaneous process that it is poorly understood, although it occurs widely in nature, since it takes place in multiple scales and involves very weak intermolecular interactions. The interests in modelling autonomous supramolecular assembly of materials is a vast challenging area in modern material science as well as in basic physical and chemical sciences. Indeed, wide varieties of supramolecular structures are possible depending on the nature of weak forces involved in complex materials made of components ranging from *micro-* to *nano-* structures. It is interesting to remark that particular spatial distributions of the supramolecular structure fluctuations have been predicted theoretically to promote quantum coherence at high temperatures.[16-19]

The SµXRD jointly with the spatial statistics allows the investigation of Levy distribution in quasi stationary states out of equilibrium,[59-61] where non-Euclidean hyperbolic space could play a key role in phenomena like chiral symmetry breaking,[62] bistability,[63] complex high temperature superconductors made of percolating pathways in filamentary hyperbolic networks[64-67], and quantum networks in living matter[2-10,68].

Furthermore, the determination of the statistical fluctuations of the myelin supramolecular structure can be important for the study of emergence of neuro degenerative processes. In fact it is possible that the onset of the neurodegenerative process associated with the transition from the functional ultrastructure to the denatured phase could be detected at an early stage by a deviation of structural fluctuations away from the Levy distribution. More specifically, both the decreasing of structural fluctuations at nanoscale and the loss of their spatial correlations constitute a measure of the degeneration degree.

Finally developing a paradigm based on fluctuations-function relation provides insights for further investigations of ultrastructure dynamics *versus* function, by studying time evolution of spatial fluctuations under external factors such as diseases and drug response.

**Materials and Methods**

**Sample preparation**

The experimental methods were carried out in accordance with the approved guidelines. Adult female frogs (Xenopus Laevis; 12 *cm* length, 180-200 *g* weight, Xenopus express, France) were housed and euthanized at the Grenoble Institute of Neurosciences with kind cooperation of Dr Andre Popov. The local committee of Grenoble Institute of Neurosciences approved the animal experimental protocol. The frogs were individually transferred in water to a separate room for euthanasia that was carried out using a terminal dose of tricaine (MS222) by immersion, terminal anaesthesia was confirmed by the absence of reflexes. Death was ensured by decapitation. Two sciatic nerves were ligated with sterile silk sutures and extracted from both thighs of freshly sacrificed Xenopus frog at approximately the same proximal-distal level through a careful dissection of the thigh. After dissection, the sciatic nerves were equilibrated in culture medium at pH 7.3 for at least 3 hours at room temperature. The culture medium was a normal Ringer's solution, containing 115 *mM* NaCl, 2.9 *mM* KCl, 1.8 *mM* $CaCl_2$, 5 *mM* HEPES (4-2-hydroxyethyl-1-piperazinyl-ethanesulfonic). Following equilibration, one of the freshly extracted nerves was immediately placed in a thin-walled quartz capillary, 1 *mm* diameter, sealed with wax and mounted perpendicular to the sample holder, for the SµXRD imaging measurements. In total we acquired 20301 X-ray diffraction patterns, for the fresh sample.

Another sciatic nerve, after dissection, was left in Petri dish and equilibrated in culture medium at pH 7.3 for 18 hours at room temperature. Following equilibration in the same described conditions, the nerve was prepared for a further SµXRD imaging session in the same day. In total we acquired 8989 X-ray diffraction patterns, for the unfresh sample.

The third sciatic nerves, after equilibration, were placed in a solution at pH 5 for 3 hours at room temperature. The culture acidic solution was an acetate buffer solution, made up by mixing 847 ml of 0.1 M acetic acid ($CH_3COOH$) and 153 ml of 0.1 M sodium acetate tri-hydrate. Following equilibration, the nerve was placed in a thin-walled quartz capillary, sealed with wax and mounted on the sample holder for a third SµXRD imaging session. In total we acquired 10201 X-ray diffraction patterns, for the denatured with pH sample.

**Experimental and data analysis**

The Scanning micro X ray Diffraction measurements of myelin of frog's sciatic nerve were performed on the ID13 beamline[26,27] of the European Synchrotron Radiation Facility, ESRF, France. The source of the synchrotron radiation beam is a 18 *mm* period in vacuum undulator. The beam is first monochromatized by a liquid nitrogen cooled Si-111 double monochromator (DMC) and then is focused by a Kirkpatrick-Baez (KB) mirror system. This optics produces an energy X-ray beam of $\lambda$=12.6 *KeV* on a 1x1 $\mu m^2$ spot. The sample holder hosts the capillary-mounted nerve with the horizontal (y) and vertical (z) translation stages with 0.1 *µm* repeatability. The sample was scanned by using a step size of 5 *µm* in both *y* and *z* direction, in order to avoid a sample damaging and autocorrelation between measured points. A Fast Readout Low Noise (FReLoN) camera (1024x1024 pixels of 100x100$\mu m^2$) is placed at a distance of 565.0 *mm* from the sample to collect the 2-D diffraction pattern in transmission. Diffraction images were calibrated using silver behenate powder ($AgC_{22}H_{43}O_2$), which has a fundamental spacing of $d_{001}$=58.38Å. We choose an exposure time of 300 *ms* for minimizing the radiation damage and keeping good photon counting statistics at the same time.[26,27] The crossed bundle is of approximately 50 myelinated axons. Therefore, the diffraction frames are an average of these axons. Considering the scale of our problem, this is an acceptable average.

We measured different regions of interest (ROIs) in the central part of the nerves around their axis to minimize the capillary geometry effect on the X-ray absorption. A typical 2-D diffraction pattern with the expected arc-rings corresponding to the Bragg diffraction orders h = 2, 3, 4, 5 is shown in **Figure 1b**. The 2-D diffraction patterns have been radially integrated to provide 1-D intensity profiles, *I(s)*, vs. transfer moment *s=2sin(θ)/λ*, after background subtraction and normalization with respect to the impinging beam (Fig. 1b). The 1-D radial profile shows the four characteristic peaks of myelin modelled with a Lorentian line shape from which we get the peak amplitude, *A(h),* and full width at half height, *w(h)*, used for the Fourier's analysis yielding electron density for each pixel according to:

$$|F(h)| = \sqrt{hA(h)w(h)} = \sqrt{hI(h)} \qquad (2)$$

for each reflection of order *h*. These structure factors, $|F(h)|$, were employed in a Fourier analysis to extract the Electron Density Distribution (EDD) of myelin:

$$\rho = \frac{F(0)}{d} + \left(\frac{2}{d}\right) \sum_{h_{min}}^{h_{max}} \pm F(h/d) \cos(2\pi rh/d) \qquad (3)$$

where the phases were taken from literature:[30,32]

$$F(h) = |F(h)|e^{i\phi_h} = \pm|F(h)| \qquad (4)$$

since the nerve is a centrosymmetric structure, so we consider just the real terms of Fourier series. The EDD obtained from the diffraction patterns measured at different sample positions in Fig. 1b, are shown in **Figure 1c**. From the differences between two adjacent maxima in the EDD profile the widths of the inter-membrane spaces at the cytoplasmic ($d_{cyt}$) and extracellular ($d_{ext}$) appositions and the thickness of the lipid bilayer ($d_{lpg}$) were obtained. From these we got the period of the structural unit[26,32] $d_\lambda=2d_{lpg}+d_{ext}+d_{cyt}$. A scheme of the myelin ultrastructure is shown in **Figure 1a.** The extrapolation of EDD at each pixel of the ROI, was performed using a customized in-house developed code written in MatLab (Mathworks, Natick, MA).

**Acknowledgments**

We are grateful to the staff of ID13 beam line at ESRF for experimental help and to Dr Andre Popov at the G.I.N for animal housing and specimen preparation. We thank T. A. Hawkins of Department of Cell and Developmental Biology at UCL London for experimental help at the early stage of the experiment. We thank superstripes-onlus for financial support. M.D.G. thanks superstripes-onlus for the RICMASS fellowship. G.C. acknowledges the support of the Institute of Crystallography of Consiglio Nazionale delle Ricerche.

**Author contributions**
All authors contributed equally to the work. M.B. provided the XRD station at ESRF; N.P. prepared the samples; M.D.G. G.C., A.B. and A.R. performed the data analysis; G.C, M.D.G., A.B., together wrote the paper.

**Additional information**
Competing financial interests: The authors declare no competing financial interests.